\documentclass[12pt]{article}
\usepackage{graphicx}
\usepackage{epsfig}
\usepackage{epstopdf}
\DeclareGraphicsExtensions{.pdf,.eps,.png,.jpg,.mps}

\usepackage{cite}

\def\lsim{\mathrel{\rlap{\lower3pt\hbox{\hskip0pt$\sim$}}
     \raise1pt\hbox{$<$}}}         %less than or approx. symbol
\def\gsim{\mathrel{\rlap{\lower4pt\hbox{\hskip1pt$\sim$}}
     \raise1pt\hbox{$>$}}}         %greater than or approx. symbol

\usepackage{amsmath}
\usepackage{amsfonts}

\begin{document}
\begin{titlepage}

\centerline{\Large \bf Notes on Fano Ratio and Portfolio Optimization}
\medskip

\centerline{Zura Kakushadze$^\S$$^\dag$\footnote{\, Zura Kakushadze, Ph.D., is the President and CEO of Quantigic$^\circledR$ Solutions LLC,
and a Full Professor at Free University of Tbilisi. Email: zura@quantigic.com} and Willie Yu$^\sharp$\footnote{\, Willie Yu, Ph.D., is a Research Fellow at Duke-NUS Medical School. Email: willie.yu@duke-nus.edu.sg}}
\bigskip

\centerline{\em $^\S$ Quantigic$^\circledR$ Solutions LLC}
\centerline{\em 1127 High Ridge Road \#135, Stamford, CT 06905\,\,\footnote{\, DISCLAIMER: This address is used by the corresponding author for no
purpose other than to indicate his professional affiliation as is customary in
publications. In particular, the contents of this paper
are not intended as an investment, legal, tax or any other such advice,
and in no way represent views of Quantigic$^\circledR$ Solutions LLC,
the website \underline{www.quantigic.com} or any of their other affiliates.
}}
\centerline{\em $^\dag$ Free University of Tbilisi, Business School \& School of Physics}
\centerline{\em 240, David Agmashenebeli Alley, Tbilisi, 0159, Georgia}
\centerline{\em $^\sharp$ Centre for Computational Biology, Duke-NUS Medical School}
\centerline{\em 8 College Road, Singapore 169857}
\medskip
\centerline{(October 9, 2017)}

\bigskip
\medskip

\begin{abstract}
{}We discuss -- in what is intended to be a pedagogical fashion -- generalized ``mean-to-risk" ratios for portfolio optimization. The Sharpe ratio is only one example of such generalized ``mean-to-risk" ratios. Another example is what we term the Fano ratio (which, unlike the Sharpe ratio, is independent of the time horizon). Thus, for long-only portfolios optimizing the Fano ratio generally results in a more diversified and less skewed portfolio (compared with optimizing the Sharpe ratio). We give an explicit algorithm for such optimization. We also discuss (Fano-ratio-inspired) long-short strategies that outperform those based on optimizing the Sharpe ratio in our backtests.
\end{abstract}
\medskip
\end{titlepage}

\newpage
\section{Introduction and Summary}

{}When constructing a (e.g., stock) portfolio, one balances risk and reward (i.e., expected return) \cite{Sharpe66}. Mean-variance optimization \cite{Markowitz} provides an implementation of this general idea. In some (somewhat limited) sense, maximizing the Sharpe ratio \cite{Sharpe66} can be taken as a justification for mean-variance optimization. Thus, without costs, bounds, constraints, etc., maximizing the Sharpe ratio of a portfolio is equivalent to mean-variance optimization. However, once, e.g., costs are included, this equivalence is gone. This begs the question:\footnote{\, Modifications of mean-variance optimization have been discussed before; see, e.g., \cite{Konno}, \cite{Rockafellar}, \cite{Bowler}, \cite{Michaud}, \cite{Braga}. E.g., in \cite{Konno} the standard deviation is replaced by MAD (that is, mean absolute deviation). Here we take a rather different approach.}

{}{\em Can we anchor portfolio optimization on quantities other than the Sharpe ratio?} In these notes we address precisely this question. The Sharpe ratio is a ratio of the (properly adjusted -- see below) expected return over the standard deviation. So, it is a ratio of the expected return to a {\em particular measure} of risk, in this case, the standard deviation. However, we can consider other measures of risk, e.g., some generic function of the variance.\footnote{\, The standard deviation is a square root of the variance, but other functions are also possible.} Thus, one property of the Sharpe ratio is that it depends on the time horizon for which it is calculated. E.g., a daily expected return and volatility give us a daily Sharpe ratio, which is on average lower (by a factor of $\sqrt{252}$, where 252 is the approximate number of trading days in a year, if we focus on stocks) than an annualized Sharpe ratio. If the daily expected return and volatility are stable in time, the Sharpe ratio goes to infinity as $\sqrt{T}$ with the time horizon $T$.

{}In contrast, the mean-to-variance ratio (i.e., the expected-return-to-variance ratio) -- which we refer to as the Fano ratio (see the next section) -- is independent of the horizon $T$ (in the aforementioned sense). We could then take the Fano ratio as the starting point for portfolio optimization. As mentioned above, more generally, we can take a ratio of the expected return to a suitable function of the variance. This is the avenue we explore in these notes, which are intended to be pedagogical.

{}In Section \ref{sec.2} we discuss maximizing generalized mean-to-risk ratios in the context of long-only portfolios. Maximizing the Fano ratio leads to simplifications (compared with the general case). Dealing with nonnegativity of the portfolio weights, just as when maximizing the Sharpe ratio, requires an iterative procedure and we provide an approximate relaxation algorithm for optimizing the Fano ratio. For long-only portfolios optimizing the Fano ratio effectively amounts to shifting the expected returns by a positive amount, which results in fewer stocks being excluded from the portfolio (including some stocks with negative expected returns), i.e., in a more diversified and less skewed portfolio (compared with optimizing the Sharpe ratio).

{}In Section \ref{sec.3} we discuss long-short portfolios, for which certain issues with optimizing the Fano ratio inspire construction of new ``multiply-optimized" strategies, which outperform optimizing the Sharpe ratio. We briefly conclude in Section \ref{sec.4}.

\newpage
\section{Generalized Mean-to-Risk Ratios}\label{sec.2}

{}Our discussion below is agnostic to the underlying tradable instruments, which a priori can be stocks, bonds, currencies, etc. However, for the sake of definiteness, let us focus on a portfolio of stocks (e.g, 2,000+ most liquid U.S. stocks). So, we have $N$ stocks with time series of (e.g., close-to-close daily, weekly, monthly or some other horizon) returns $R_{is}$, $i=1,\dots,N$. Here the index $s=1,\dots,T$ labels trading days on which these returns are computed ($s=1$ labels the most recent date).\footnote{\, Here the returns $R_{is}$ are defined as excess returns w.r.t. a risk-free return. In the case of dollar-neutral portfolios this is not crucial. However, here we do not require dollar neutrality.}

{}Above, $R_{is}$ are the realized returns (ex-post). We can also define expected returns (ex-ante) via, e.g., moving averages:
\begin{equation}\label{eta}
 E_{is} = {1\over d} \sum_{s^\prime = s + 1}^{s + d} R_{is^\prime}
\end{equation}
Thus, if $R_{is}$ are daily returns, then $E_{is}$ are $d$-day moving averages. We emphasize that (\ref{eta}) is only an example and there are myriad other ways of constructing $E_{is}$. Generally, expected returns can be quite convoluted and have no simple financial interpretation, e.g., machine learning based expected returns \cite{101}. In the following, for the sake of simplicity, we will omit the index $s$ and refer to expected returns as $E_i$. Thus, we can think of $E_i$ as the expected returns for $s=1$ (i.e., ``today's" date). What is important is that $E_i$ are computed out-of-sample.

{}Next, we can define a sample covariance matrix $C_{ij}$ based on the time series $E_{is}$ or $R_{is}$ and also computed out-of-sample (ex-ante). In what follows it may appear natural to compute $C_{ij}$ based on the expected returns $E_{is}$ as opposed to the realized returns $R_{is}$. However, in practice, in many cases it can be (much) simpler to compute $C_{ij}$ based on $R_{is}$. In some cases basing $C_{ij}$ on $E_{is}$ may not even be practicable. One issue is that typically the lookback -- i.e., the number of datapoints in the time series, call it $M$ -- is insufficient to compute $C_{ij}$ reliably. Thus, if $M < N + 1$, then the sample covariance matrix $C_{ij}$ is singular, whereas for our purposes below $C_{ij}$ must be positive-definite. Furthermore, unless $M \gg N$, which is rarely if ever the case in practice, the off-diagonal elements (in particular, the correlations -- the diagonal elements are relatively stable) are highly unstable out-of-sample rendering $C_{ij}$ essentially useless (unpredictive out-of-sample). So, in practice one replaces the sample covariance matrix $C_{ij}$ via a model covariance matrix, call it $\Gamma_{ij}$, such as a multifactor risk model.\footnote{\, For a general discussion, see, e.g., \cite{GK}. For an explicit open-source implementation of a general multifactor risk model for equities, see \cite{HetPlus}.} If built in-house, $\Gamma_{ij}$ could a priori be built based on $E_{is}$ (among other things). If it is a third-party product, naturally, it is built based on $R_{is}$ (or some other returns). In any event, here we will not delve into how $\Gamma_{ij}$ is built. We will simply assume that $C_{ij}$ below is identified with some model covariance matrix $\Gamma_{ij}$, which is i) positive-definite and ii) sufficiently stable out-of-sample.

\subsection{Generalized Mean-to-Risk Ratios for Portfolios}

{}Now we can define portfolio risk. Let us assume that our portfolio consists of our $N$ stocks with weights $w_i$. A priori some of these weights can be 0 or negative. The normalization condition for the weights is
\begin{equation}\label{norm}
 \sum_{i=1}^N |w_i| = 1
\end{equation}
Below we will consider a case with nonnegative weights; for now $w_i$ are general.

{}The expected return of the portfolio is given by
\begin{equation}\label{E}
 E = \sum_{i=1}^N w_i~E_i
\end{equation}
The expected variance of the portfolio is given by
\begin{equation}\label{V}
 V = \sum_{i,j=1}^N C_{ij}~w_i~w_j
\end{equation}
We can define the Sharpe ratio \cite{Sharpe94} of the portfolio via
\begin{equation}\label{SR}
 S = {E \over \sqrt{V}}
\end{equation}
A nice thing about the Sharpe ratio is that it is invariant under the formal rescalings $w_i\rightarrow \zeta~ w_i$, where $\zeta > 0$. This rescaling invariance is the reason why in the absence of trading costs, bounds, etc., maximizing the Sharpe ratio is equivalent to mean-variance optimization \cite{Markowitz}.\footnote{\, More precisely, there is a single exception to this, which is the case of linear costs for establishing trades -- see \cite{MeanRev} for details.} Indeed, maximizing (\ref{SR}) (i.e., we find the maximum of $S$ w.r.t. $w_i$)\footnote{\, Which can be done by ignoring (\ref{norm}) due to the aforesaid rescaling invariance as we can always rescale the weights obtained via such maximization to conform to (\ref{norm}). In fact, maximizing (\ref{SR}) fixes $w_i$ only up to an overall normalization factor.} is equivalent to maximizing (w.r.t. $w_i$) the objective function
\begin{equation}
 g = E - {\lambda\over 2}~V
\end{equation}
and $\lambda$ is fixed (after maximization) via (\ref{norm}). So, in some (somewhat limited) sense, maximizing the Sharpe ratio can be taken as a justification for mean-variance optimization. However, once, e.g., trading costs, etc., are added, maximizing the Sharpe ratio is no longer equivalent to mean-variance optimization \cite{MeanRev}.

{}Furthermore, one property of the Sharpe ratio is that it depends on the time horizon for which it is calculated. E.g., a daily expected return and volatility give us a daily Sharpe ratio, which is on average lower (by a factor of $\sqrt{d}$, where $d\approx 252$ is the approximate number of trading days in a year, if we focus on stocks) than an annualized Sharpe ratio. If the daily expected return and volatility do not change much in time, then the Sharpe ratio goes to infinity as $\sqrt{T}$ with the time horizon $T$. So, a practical way of thinking about the Sharpe ratio is that, if, say, the annualized Sharpe ratio is 2, then the probability of losing money in a given year is less than about 2.3\% (assuming normally distributed realized returns, that is, which can be farfetched -- see below).\footnote{\, Recall the 68-95-99.7 rule: if $x$ is a normally distributed variable with mean $\mu$ and standard deviation $\sigma$, then we have the following probabilities: $\mbox{Pr}(\mu - n\sigma \leq x \leq \mu + n\sigma) = P_n$, $P_1\approx 68.27\%$, $P_2 \approx 95.45\%$, $P_3\approx 99.73\%$. The probability of losing money when the Sharpe ratio equals $n$ (i.e., $\mu = n\sigma$) then is ${\widetilde P} = (1 - P_n)/2$. So, we have ${\widetilde P}_1 \approx 15.9\%$, ${\widetilde P}_2 \approx 2.3\%$, ${\widetilde P}_3 \approx 0.14\%$. However, this does not take into account leverage, margin calls, investor withdrawals and other such nuances.} Can we define a ratio independent of the time horizon?\footnote{\, Another known issue with the Sharpe ratio maximization and mean-variance optimization is that one can get portfolios with low degree of diversification. E.g., consider a simple example where all $E_i > 0$ and the matrix $C_{ij} = \sigma_i^2~\delta_{ij}$ is diagonal (uncorrelated returns -- this is not a crucial assumption here, the following can happen even for correlated returns). The weights that maximize the Sharpe ratio are given by $w_i = \gamma E_i/\sigma_i^2$, where $\gamma$ is fixed via (\ref{norm}). Now consider a case where all $E_i$ are small except for one. Then we can have all weights but one small and most of the investment will be allocated to the corresponding single stock thereby forgoing diversification.}

{}The answer is affirmative. We can simply define the mean-to-variance ratio:\footnote{\, A mean-to-variance ratio test is advocated as a complement to a coefficient of variation test and a Sharpe ratio test in \cite{Bai1}. Also see \cite{Bai2}.}
\begin{equation}\label{FR}
 F = {E\over V}
\end{equation}
This ratio is independent of the horizon $T$ (in the aforementioned sense). Other than tautological ``mean-to-variance ratio", this ratio apparently has not been named in finance. It would appear appropriate to term it {\em the Fano ratio} due to its relationship to the Fano factor \cite{Fano} named after Ugo Fano, an Italian American physicist. The Fano factor (a.k.a. variance-to-mean ratio and index of dispersion), in our notations, is simply $V/E$, i.e., the inverse of the Fano ratio (\ref{FR}).\footnote{\, Similarly, the Sharpe ratio is the inverse of the coefficient of variation $\sqrt{V}/E$.} Up to a factor of 2, it is the same as the ratio
\begin{equation}
 \kappa = {2E\over V}
\end{equation}
discussed in \cite{Bubbles} in the context of stock price bubbles, where it was argued that the dimensionless ratio $\kappa$ can be used to define a criterion for when a stock (or a similar instrument) is not a good investment in the long term, which can happen even if the expected return is positive. Thus, assuming log-normal distribution for stock prices, this criterion (i.e., that the stock is not a good investment in a long run) is
\begin{equation}
 \kappa < 1
\end{equation}
This criterion for the Fano ratio (defined for a single stock) would be $F < 1/2$.

{}So, naturally, we can ask: why not maximize the Fano ratio (instead of the Sharpe ratio)? In fact, we can define more general ``mean-to-risk" ratios via
\begin{equation}\label{GR}
 G = {E \over f(V)}
\end{equation}
where $f(V)$ is some function of $V$. We can then maximize $G$ instead of $S$ (or $F$).

{}Recall, however, that we could ignore the normalization condition (\ref{norm}) when maximizing the Sharpe ratio as the latter is invariant under the rescalings $w_i\rightarrow \zeta~w_i$. Such invariance is gone in the case of the Fano ratio or more general ratios $G$ defined via (\ref{GR}). So, the maximization problem becomes more nontrivial. Thus, we must maximize the objective function
\begin{equation}\label{g1}
 g = G + \mu\left(\sum_{i=1}^N |w_i| - 1\right)
\end{equation}
where $\mu$ is a Lagrange multiplier. The modulus complicates things quite a bit (see Section \ref{sec.3}). Therefore, for the sake of simplicity, for now let us focus on the case of long-only portfolios, where $w_i\geq 0$. Then our objective function simplifies:
\begin{equation}\label{g}
 g = G + \mu\left(\sum_{i=1}^N w_i - 1\right)
\end{equation}
However, now we have bounds
\begin{equation}\label{bounds}
 w_i \geq 0
\end{equation}
To get a flavor of the problem at hand, at first we will ignore the bounds (\ref{bounds}) when solving the maximization problem and then incorporate them via a certain trick.

\subsection{Maximization Ignoring Bounds}

{}Maximizing (\ref{g}) w.r.t. $w_i$ and $\mu$ (and ignoring the bounds (\ref{bounds})), we get the following solution ($f^\prime(V)$ is the first derivative w.r.t. $V$, and $C^{-1}_{ij}$ is the inverse of $C_{ij}$):
\begin{eqnarray}
 &&w_i = a~{\widehat \eta}_i + b~{\widehat \nu_i}\label{cc1}\\
 &&{\widehat \eta}_i = \sum_{j=1}^N C^{-1}_{ij}~E_j\\
 &&{\widehat \nu}_i = \sum_{j=1}^N C^{-1}_{ij}~\nu_j\\
 &&a = {f(V)\over 2Ef^\prime(V)}\label{cc3}\\
 &&b = V - {f(V)\over 2f^\prime(V)}\label{cc4}
\end{eqnarray}
Here $\nu_i \equiv 1$ is the unit $N$-vector (using which might appear redundant at first, but will be useful later). So, we have three unknowns, $E$, $V$ and $a$. Using the condition $\sum_{i=1}^N w_i = 1$ and the definitions (\ref{E}) and (\ref{V}), we have the following equations:\footnote{\, Note that (\ref{c2}) follows from (\ref{cc3}), (\ref{cc4}), (\ref{c1}) and (\ref{c3}).}
\begin{eqnarray}
 &&a\gamma + b\beta^2 = 1\label{c1}\\
 &&E = a\alpha^2 + b\gamma\label{c2}\\
 &&V = a^2\alpha^2 + 2ab\gamma + b^2\beta^2 = a^2\alpha^2 + (a\gamma + 1)b = a^2\alpha^2 + (1 - a^2\gamma^2) / \beta^2\label{c3}
\end{eqnarray}
where
\begin{eqnarray}
 &&\alpha^2 = \sum_{i,j = 1}^N C^{-1}_{ij}~E_i~E_j\\
 &&\beta^2 = \sum_{i,j = 1}^N C^{-1}_{ij}~\nu_i~\nu_j\\
 &&\gamma = \sum_{i,j = 1}^N C^{-1}_{ij}~E_i~\nu_j
\end{eqnarray}
and in (\ref{c3}) we repeatedly used (\ref{c1}). So we can express $a$ via $V$:
\begin{equation}\label{aa}
 a^2 = {{V\beta^2 - 1}\over {\alpha^2\beta^2 - \gamma^2}}
\end{equation}
Combining (\ref{c1}), (\ref{cc4}) and (\ref{aa}), we get the following equation involving $V$ only:
\begin{equation}\label{quad}
 \gamma^2~{{V\beta^2 - 1}\over {\alpha^2\beta^2 - \gamma^2}} = \left(1 + \beta^2\left[{f(V)\over 2f^\prime(V)} - V\right]\right)^2
\end{equation}
For general $f(V)$ this equation is transcendental. In some cases it simplifies.

{}Let us start with the case of the Sharpe ratio $f(V) = \sqrt{V}$. Then we have the familiar solution $b = 0$ and $a = 1/\gamma$. If $f(V) = V^p$, where $p > 0$ and $p\neq 1/2$, then (\ref{quad}) is a quadratic equation for $V$ and can be readily solved. It is also a quadratic equation when $f(V) = \exp(\xi V)$. For
$f(V) = \exp(\xi V^2)$ the equation is quartic.

\subsubsection{Maximizing Fano Ratio}

{}While the aforesaid quadratic equations can be solved, generally they involve radicals and are not particularly illuminating. However, in the case of the Fano ratio, i.e., when $f(V) = V$, things further simplify and there are no radicals. The solution is:
\begin{eqnarray}\label{aaa}
 && a = {1\over{\alpha\beta + \gamma}}\\
 && b = {\alpha\over{\beta(\alpha\beta + \gamma)}}
\end{eqnarray}
and we have
\begin{eqnarray}
 && E = {\alpha \over \beta}\label{E.full}\\
 && V = {2\alpha\over{\beta(\alpha\beta + \gamma)}}\\
 && F = {1\over 2}\left(\alpha\beta + \gamma\right)\label{FF1}\\
 && S = \sqrt{\alpha(\alpha\beta+\gamma)\over 2\beta}
\end{eqnarray}
Had we maximized the Sharpe ratio, we would get $E_* = \alpha^2/\gamma$, $V_* = \alpha^2/\gamma^2$, $F_* = \gamma$ and $S_* = \alpha$. Since $\gamma < \alpha\beta$, it follows that $S < S_*$ and $F > F_*$ (which should come as no surprise as $F$ and $S_*$ are the maximum possible values thereof), and $E < E_*$ and $V < V_*$. So, maximizing the Fano ratio produces a portfolio with a lower expected return but also a lower expected volatility than maximizing the Sharpe ratio.\footnote{\, Note that in portfolios based on maximizing the Sharpe ratio the realized expected return and Sharpe ratio can be vastly different from their expected values based on optimization. Therefore, the fact that the expected return is higher when we maximize the Sharpe ratio as compared to when we maximize the Fano ratio means little in terms of what the realized return will be.}

\subsection{Incorporating Bounds}

{}The solution (\ref{cc1}) is not necessarily good in the sense that some $w_i$ might be negative. Indeed, even if all $E_i$ are nonnegative, we can have negative $w_i$ due to the off-diagonal elements in $C_{ij}$. So, we must incorporate the bounds (\ref{bounds}) into the solution somehow.

{}The issue is that we are not dealing with a quadratic optimization problem here.\footnote{\, This is also the case when maximizing the Sharpe ratio in the presence of bounds.} However, not all is lost and the following trick provides a reasonable approximation. Thus, the solution (\ref{cc1}) {\em formally} can be thought of as the solution to maximizing the following quadratic objective function ($\lambda$ is fixed after solving for $w_i$ by rescaling them such that $\sum_{i=1}^N w_i = 1$, which rescaling is not affected by the bounds (\ref{bounds}))
\begin{equation}\label{gt}
 {\widetilde g} = {\widetilde E} - {\lambda\over 2}~V
\end{equation}
subject to the bounds (\ref{bounds}), where (note that $\lambda = \alpha(J)\beta(J)+\gamma(J)$)
\begin{eqnarray}
 &&{\widetilde E} = \sum_{i=1}^N w_i~{\widetilde E}_i\\
 &&{\widetilde E}_i = E_i + E(J)~\nu_i\label{E-eff-long}\\
 &&E(J) = {\alpha(J)\over\beta(J)}
\end{eqnarray}
\begin{eqnarray}
 &&[\alpha(J)]^2 = \sum_{i,j \in J} [C(J)]^{-1}_{ij}~E_i~E_j\\
 &&[\beta(J)]^2 = \sum_{i,j \in J} [C(J)]^{-1}_{ij}~\nu_i~\nu_j\\
 &&\gamma(J) = \sum_{i,j \in J} [C(J)]^{-1}_{ij}~E_i~\nu_j
\end{eqnarray}
where $J = \{i|w_i > 0\}$ is the subset of positive weights, $[C(J)]^{-1}_{ij}$ is the $N(J) \times N(J)$ matrix inverse to the $N(J) \times N(J)$ matrix $[C(J)]_{ij}$ obtained from $C_{ij}$ by restricting $i,j\in J$. (Here $N(J) = |J|$ is the number of elements in $J$.) So, the catch is that $E(J)$, and thereby ${\widetilde E}_i$, depend on $J$, which is unknown. Had $E(J)$ been known a priori, then we would simply minimize (\ref{gt}) subject to the bounds (\ref{bounds}) via standard quadratic optimization techniques.\footnote{\, See, e.g., \cite{DG}, \cite{dH}, \cite{Jansen}, \cite{MeanRev}, \cite{Murty}, \cite{Pang}, and references therein.} So, here is a relaxation algorithm that {\em approximates} the optimal solution. At the initial iteration, we assume that $J^{(0)}$ is the full set $\{1,\dots,N\}$ and compute $w_i$ via (\ref{cc1}). If all $w_i\geq 0$, then there is nothing else to do, we are done. So, let us assume that the set ${\widetilde J}^{(0)} = \{i| w_i < 0\}$ is not empty. Let us take the value of $\ell \in {\widetilde J}^{(0)}$ for which $F_\ell = \min(F_i)$, where $F_i = E_i/C_{ii}$ are the Fano ratios for each stock.\footnote{\, If there are multiple values of $\ell$ for which $F_\ell = \min(F_i)$, then we take the value of $\ell$ for which $C_{\ell\ell} = \max(C_{ii})$, and if there are still multiple values of $\ell$ remaining, we simply take the lowest $\ell$.\label{fn.ell}} We then permanently set $w_\ell = 0$, take $J^{(1)} = J^{(0)}\setminus\ell$ and compute $w_i$ via (\ref{cc1}). If the resulting $w_i\geq 0$ for all $i\in J^{(1)}$, then we are done. So, let us assume that the set ${\widetilde J}^{(1)} = \{i| w_i < 0, i\in J^{(1)}\}$ is not empty. Let us take the value of $\ell^\prime \in {\widetilde J}^{(1)}$ for which $F_{\ell^\prime} = \min(F_i)$, $i\in J^{(1)}$ (see fn.\ref{fn.ell}). We then permanently set $w_{\ell^\prime} = 0$, take $J^{(2)} = J^{(1)}\setminus\ell^\prime$ and compute $w_i$ via (\ref{cc1}). And so on. We repeat this procedure until at some $k$-th iteration all $w_i\geq 0$ for $i \in J^{(k)}$. As always, one issue with this relaxation algorithm is the computational cost: we must compute the inverse matrix $[C(J)]^{-1}_{ij}$ at each iteration. However, for a $K$-factor model of the form (here $\xi_i^2$ is the specific a.k.a. idiosyncratic risk, $\Omega_{iA}$, $A=1,\dots,K$ is the factor loadings matrix, and $\phi_{AB}$ is the factor covariance matrix)
\begin{equation}\label{factor}
 C_{ij} = \xi_i^2~\delta_{ij} + \sum_{A,B=1}^K \Omega_{iA}~\phi_{AB}~\Omega_{jB}
\end{equation}
to compute $[C(J)]^{-1}_{ij}$, we only need to invert the $K\times K$ matrix $\phi_{AB}$ once, plus we must invert a $K\times K$ matrix\footnote{\, Iteratively dropping the stock with the lowest Fano ratio is only an approximation. However, it is computationally feasible. E.g., iteratively dropping the stock with the smallest impact on the full Fano ratio (\ref{FF1}) would prohibitively require inverting $\sim N$ matrices (\ref{QJ}) at each iteration.}
\begin{equation}\label{QJ}
 [Q(J)]_{AB} = \phi^{-1}_{AB} + \sum_{i\in J} {1\over\xi_i^2}~\Omega_{iA}~\Omega_{iB}
\end{equation}
at each iteration. However, these inversions are much cheaper assuming $K \ll N$.

\subsection{The ``Market" Mode}

{}The issue we wish to address next is equally pertinent to maximizing the Sharpe ratio, the Fano ratio and the generalized mean-to-risk ratios we discuss above. For the sake of definiteness and simplicity, let us focus on the case of maximizing the Sharpe ratio. Let us ignore the bounds (\ref{bounds}) for a moment. Then the weights are given by
\begin{eqnarray}
 && w_i = a~\sum_{j = 1}^N C^{-1}_{ij}~E_j\\
 && a^{-1} = \sum_{i,j = 1}^N C^{-1}_{ij}~E_i~\nu_j
\end{eqnarray}
For a typical configuration, even if all $E_i$ are nonnegative, close to 50\% of the weights $w_i$ can be negative. Thus, to illustrate this point, let us consider the following ``toy" covariance matrix:\footnote{\, This is an example of a 1-factor model.} $C_{ij} = \sigma_i\sigma_j\Psi_{ij}$, where $\sigma_i^2$ are the variances; the correlation matrix $\Psi_{ij} = \left(1-\rho\right)\delta_{ij} + \rho \nu_i \nu_j$; and $\nu_i\equiv 1$ is the unit $N$-vector. I.e., all $N$ stocks have uniform pair-wise correlations equal $\rho$. Inverting this matrix gives the following weights:
\begin{equation}\label{SR.w}
 w_i = {a\over\sigma_i\left(1-\rho\right)}\left[{\widetilde E}_i - {\rho~\nu_i\over{1 + (N-1)\rho}}~\sum_{j=1}^N {\widetilde E}_j\right]
\end{equation}
where ${\widetilde E}_i = E_i/\sigma_i$ are the normalized expected returns. Generically, the latter are expected to be roughly symmetrically distributed around their mean. It is then evident from (\ref{SR.w}) that, unless $\rho \ll 1/N$, roughly 50\% of the weights $w_i$ are negative.

{}Now, in practice the correlation matrix with uniform pair-wise correlations is unrealistic. However, the above issue persists even for realistic correlation matrices. Thus, consider a general correlation matrix $\Psi_{ij}$. We can always write it as
\begin{equation}
 \Psi_{ij} = \left(1 - \rho\right)\delta_{ij} + \rho\nu_i\nu_j + \Delta_{ij} = \Psi^\prime + \Delta_{ij}
\end{equation}
Here $\rho = {1\over N(N-1)}\sum_{i,j=1;~i\neq j}^N\Psi_{ij}$ is the {\em average} pair-wise correlation, and $\sum_{i,j=1}^N \Delta_{ij} = 0$. In the zeroth approximation we can drop $\Delta_{ij}$, i.e., $\Psi_{ij}\approx \Psi^\prime_{ij}$. Its first principal component $U^{(1)}_i = \nu_i/\sqrt{N}$. It describes the ``market" mode \cite{CFM},\footnote{\, Also see \cite{Billion}.} i.e., the average correlation of all stocks, which is nonzero (and not small, definitely $\rho\not\ll 1/N$).\footnote{\, Note that the eigenvalue of $\Psi^\prime_{ij}$ corresponding to $U^{(1)}_i$ is $\lambda^\prime_* = 1 + \rho \left(N-1\right)$.} The ``market" mode corresponds to the overall movement of the broad market, which affects all stocks (to varying degrees) -- cash inflow (outflow) into (from) the market tends to push stock prices higher (lower). This is the market risk factor. To mitigate this risk factor, one can, e.g., hold a dollar-neutral portfolio of stocks. However, long-only portfolios are exposed to market risk by construction.

{}So, neutralizing the market risk factor while maximizing the Sharpe ratio is an unwelcome feature. Why? Because at the end we impose the bounds (\ref{bounds}) anyway, so the market risk is still present, but the resultant portfolio gets artificially distorted due to pushing the negative weights (that is, in the unbounded optimization) to zero thereby also affecting the positive weights. The culprit here is that, when maximizing the Sharpe ratio using a covariance matrix that includes the ``market" mode, we approximately neutralize the portfolio w.r.t. the market risk. Put differently, we hedge against all stocks -- i.e., the broad market -- going bust. However, holding a long-only portfolio with thousands of stocks invariably is exposed to the broad market. So, we must eliminate the ``market" mode out of the covariance matrix.\footnote{\, Here one can argue that one can build a ``market-neutral" long-only portfolio by picking stock weights such that they are neutral w.r.t. market betas by utilizing the fact that some betas can be negative. However, not only do the betas tend to be highly unstable out-of-sample, we still have the bounds (\ref{bounds}) (which are not that easy to satisfy for beta-neutral portfolios), and neutralizing against the ``market" mode (whose elements are all positive) in no way is helpful in building a beta-neutral long-only portfolio. For a review of some market-neutral strategies (which, however, are not long-only), see, e.g., \cite{Lo} and references therein.}

{}In the context of factor models (\ref{factor}) this can be achieved relatively easily. In this case (ignoring the bounds (\ref{bounds})) we have (as above, $a$ is fixed via $\sum_{i=1}^N w_i = 1$)
\begin{eqnarray}
 && w_i = a~\left[{E_i \over \xi_i^2} - {1 \over \xi_i^2}~\sum_{A,B=1}^K \Omega_{iA}~Q^{-1}_{AB}~\sum_{j=1}^N {E_j \over \xi_j^2}~\Omega_{jB}\right]\\
 && Q_{AB} = \phi^{-1}_{AB} + \sum_{i=1}^N {1\over\xi_i^2}~\Omega_{iA}~\Omega_{iB}
\end{eqnarray}
Eliminating the ``market" mode from $C_{ij}$ then amounts to requiring that the factor loadings matrix is orthogonal to some {\em positive} $N$-vector $v_i > 0$:
\begin{equation}
 \sum_{i=1}^N v_i~\Omega_{iA}\equiv 0
\end{equation}
Then we no longer have roughly 50\% of negative weights $w_i$. While some of these weights can still be negative, typically the number of such negative weights will be relatively small compared with $N$ (assuming all $E_i\geq 0$, that is). So, what are $v_i$?

{}One -- but not the only -- way of thinking about $v_i$ is that they are the weights of some {\em benchmark} long-only portfolio: $v_i = w^{benchmark}_i > 0$. The choice of this benchmark portfolio is not all that critical provided it is reasonably diversified. For example, we can take\footnote{\, Up to an overall normalization, that is.} $v_i \equiv 1$, i.e., an equally-weighted benchmark. We can take $v_i = 1/\sigma_i^2$ or $v_i = 1/\xi_i^2$. This can skew the portfolio toward low-volatility (which are typically large market cap) stocks. To mitigate this, we can Windsorize or otherwise deal with the tails in the skewed (roughly log-normal) distribution of $\sigma_i$ (or $\xi_i$). Etc.

\subsection{Statistical Risk Models}

{}Statistical risk models \cite{StatRM} provide a particularly simple example of factor models, where the factor covariance matrix is diagonal. Thus, let $\Psi^{sample}_{ij}$ be the sample correlation matrix computed based on time series of historical returns. $\Psi^{sample}_{ij}$ can be singular. This will not affect our discussion below. The sample covariance matrix is $C^{sample}_{ij} = \sigma_i\sigma_j\Psi^{sample}_{ij}$. We can construct a statistical risk model covariance matrix $C_{ij}$ as follows:
\begin{eqnarray}
 &&C_{ij} = \sigma_i\sigma_j\Psi_{ij}\\
 &&\Psi_{ij} = {\widetilde \xi}_i^2~\delta_{ij} + \sum_{a=1}^K \lambda^{(a)}~V^{(a)}_i~V^{(a)}_j\\
 &&{\widetilde\xi}_i^2 = 1 - \sum_{a=K+1}^r \lambda^{(a)}~[V^{(a)}_i]^2
\end{eqnarray}
Here $V^{(a)}_i$ are the principal components of the matrix $\Psi^{sample}_{ij}$ with the corresponding eigenvalues in the descending order: $\lambda^{(1)} > \lambda^{(2)} > \dots > \lambda^{(r)}$, where $r$ is the rank of $\Psi^{sample}_{ij}$ (if $r < N$, then for $a > r$ we have $\lambda^{(a)} = 0$). The number of factors $K$ is determined via (truncated or rounded) eRank (effective rank) \cite{RV} -- see \cite{StatRM} for details. The issue with the so-constructed $C_{ij}$ is that it contains the ``market" mode. Indeed, without loss of generality we can assume that all elements of the first principal component $V^{(1)}_i > 0$ -- this can be ensured by, if need be, changing the basis as follows: $C_{ij} \rightarrow \epsilon_i\epsilon_j C_{ij}$, where $\epsilon_i = \mbox{sign}(V^{(1)}_i)$. Then the all-positive $V^{(1)}_i$ can be regarded as the ``market" mode \cite{CFM}. In fact, for large $N$ we have $V^{(1)}_i \approx 1/\sqrt{N}$. Note that higher principal components $V^{(a > 1)}_i$ invariably have negative elements. So, we need to eliminate the first principal component. This can be achieved simply by defining
\begin{eqnarray}
 &&C_{ij} = \sigma_i\sigma_j{\widehat \Psi}_{ij}\\
 &&{\widehat \Psi}_{ij} = {\widehat \xi}_i^2~\delta_{ij} + \sum_{a=2}^K \lambda^{(a)}~V^{(a)}_i~V^{(a)}_j\\
 &&{\widehat\xi}_i^2 = 1 - \lambda^{(1)}~[V^{(1)}_i]^2 - \sum_{a=K+1}^r \lambda^{(a)}~[V^{(a)}_i]^2
\end{eqnarray}
This is not the only possible definition, but it is as good as any other. With this definition we can think of the benchmark portfolio as that with $v_i = V^{(1)}_i / \sigma_i$.

{}Our discussion above is for maximizing the Sharpe ratio but equally applies to maximizing the Fano ratio and the generalized mean-to-risk ratios. This is because in all these cases the weights involve inverting the covariance matrix. Indeed, in (\ref{cc1}) we have $w_i = a \sum_{j=1}^N C^{-1}_{ij}~E^\prime_j$, where $E^\prime_i = E_i + \nu_i~b/a$, so the above still applies.

\subsection{Why Is This Useful?}\label{sub.why}

{}For long-only portfolios optimizing the Fano ratio {\em effectively} amounts to shifting the expected returns by a positive amount via (\ref{E-eff-long}), which results in fewer stocks being excluded from the portfolio due to the bounds (\ref{bounds}) (including some stocks with negative expected returns $E_i$, for which the effective returns ${\widetilde E}_i$ can be positive), i.e., in a more diversified and less skewed portfolio (compared with optimizing the Sharpe ratio). This is evident for a diagonal matrix $C_{ij} = \sigma_i^2~\delta_{ij}$. And this conclusion persists for non-diagonal $C_{ij}$ of the factor model form with the ``market" mode removed. This can be illustrated using a simple 1-factor model of the form $C_{ij} = \sigma_i\sigma_j\Psi_{ij}$, where the correlation matrix $\Psi_{ij} = \left(1-\rho\right)\delta_{ij} + \rho s_i s_j$, and $s_i = 1$ for half of the values of $i$, and $s_i=-1$ for the other half (the number of stocks $N$ is assumed to be even). (As above, let us ignore the bounds for a moment.) Then we have (here $a$ is given by (\ref{aaa}))
\begin{eqnarray}\label{w.uni}
 &&w_i = {a\over\sigma_i\left(1-\rho\right)}\left[{\widetilde E}^\prime_i - {\rho~s_i\over{1 + (N-1)\rho}}~\sum_{j = 1}^N {\widetilde E}^\prime_j~s_j\right]\\
 &&{\widetilde E}^\prime_i = {\widetilde E}_i + {\alpha\over\beta}~{\nu_i\over\sigma_i}\\
 &&{\widetilde E}_i = E_i / \sigma_i\\
 &&\alpha^2 = {1\over\left(1-\rho\right)}\left[\sum_{i=1}^N {\widetilde E}^2_i - {\rho\over{1 + (N-1)\rho}}~\left(\sum_{i = 1}^N {\widetilde E}_i~s_i\right)^2\right]\\
 &&\beta^2 = {1\over\left(1-\rho\right)}\left[\sum_{i=1}^N {1\over\sigma_i^2} - {\rho\over{1 + (N-1)\rho}}~\left(\sum_{i = 1}^N {s_i\over\sigma_i}\right)^2\right]
\end{eqnarray}
It is reasonable to assume that there is no substantial correlation between the values of $\sigma_i$ and the signs $s_i$, or the values of $E_i$ and $s_i$. Then we can estimate that $\left|\sum_{i=1}^N s_i/\sigma_i\right|\lsim \sqrt{N}/\sigma_*$, where $N/\sigma_*^2 = \sum_{i=1}^N 1/\sigma_i^2$. Similarly, $\left|\sum_{i=1}^N {\widetilde E}_i~s_i \right|\lsim \sqrt{N}~{\widetilde E}_*$, where $N~{\widetilde E}_*^2 = \sum_{i=1}^N {\widetilde E}_i^2$.
Further, we can reasonably assume that $\rho\not\ll 1/N$ (and $N\gg 1$). Then we have
\begin{eqnarray}
 &&\alpha^2 = {N{\widetilde E}_*^2\over\left(1-\rho\right)}\left[1 - {\cal O}(1/N)\right]\\
 &&\beta^2 = {N\over\sigma_*^2\left(1-\rho\right)}\left[1 - {\cal O}(1/N)\right]
\end{eqnarray}
and, up to terms suppressed by $1/N$, the weights are given by
\begin{equation}\label{w.fano.1-factor}
 w_i \approx {a\over\sigma_i\left(1-\rho\right)}\left[{\widetilde E}_i + {\widetilde E}_*~\nu_i~{\sigma_*\over\sigma_i}  - {s_i\over N}\sum_{j = 1}^N \left({\widetilde E}_j~s_j + {\widetilde E}_*~{\sigma_*~s_j\over\sigma_j}\right)\right]
\end{equation}
In this expression the terms containing ${\widetilde E}_*$ are pertinent to optimizing the Fano ratio; the other two terms are present when optimizing the Sharpe ratio (in which case the overall normalization coefficient $a$ is different). And it is precisely the second term in the square brackets in (\ref{w.fano.1-factor}) that makes the difference here. Here is why and how.

{}Based on our argument above, the term in (\ref{w.fano.1-factor}) containing a sum over $j$ has a magnitude of order ${\widetilde E}_*/\sqrt{N}$. Now consider the values of the index $i$ such that $\sigma_i\ll\sqrt{N}~\sigma_*$. For such $i$ the second term in (\ref{w.fano.1-factor}) dominates the term containing the sum and we have
\begin{equation}
 w_i \approx {a\over\sigma_i\left(1-\rho\right)}\left[{\widetilde E}_i + {\widetilde E}_*~\nu_i~{\sigma_*\over\sigma_i}\right]
\end{equation}
For such values of $i$, $w_i$ can be positive even for negative returns $E_i$. This is because i) the contributions of the off-diagonal terms in the covariance matrix $C_{ij}$ into the optimization are suppressed for such $i$, and ii) the intrinsic-to-Fano-ratio term (proportional to ${\widetilde E}_*$) provides an additive positive contribution. This reduces the number of stocks with negative weights (when we ignore the bounds, that is), which are then ``pushed up" when we include the bounds. And this additive contribution is positive even for the values of $i$ for which $\sigma_i\not\ll\sqrt{N}~\sigma_*$. Let us quantify this.

\begin{figure}[ht]
\centering
\includegraphics[scale=1.0]{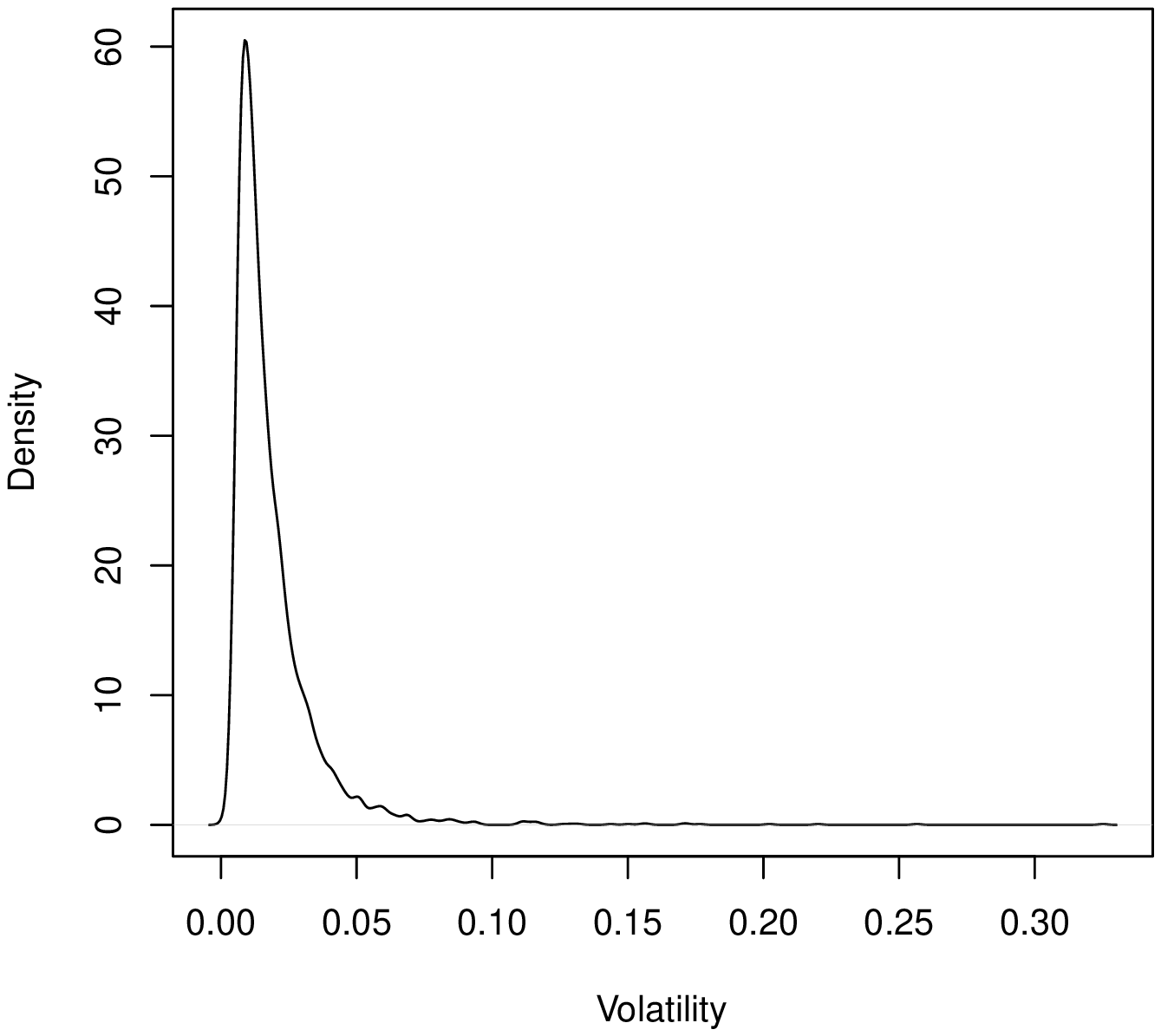}
\caption{Density of volatility for the universe of 3810 stocks (see Section \ref{sub.why}).
}
\label{Fig1}
\end{figure}

\begin{figure}[ht]
\centering
\includegraphics[scale=1.0]{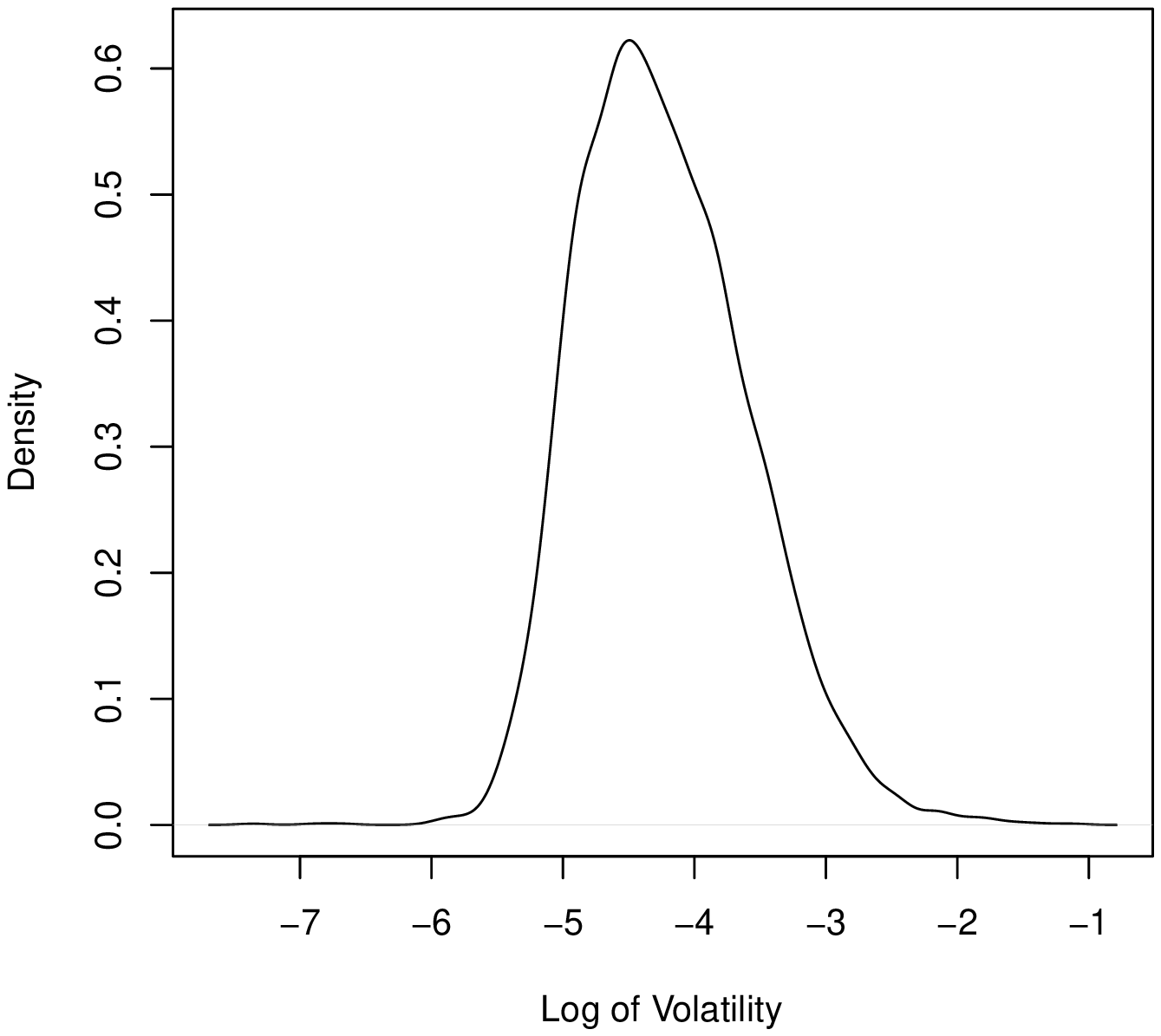}
\caption{Density of log-volatility for the universe of 3810 stocks (see Section \ref{sub.why}).
}
\label{Fig2}
\end{figure}

\begin{figure}[ht]
\centering
\includegraphics[scale=1.0]{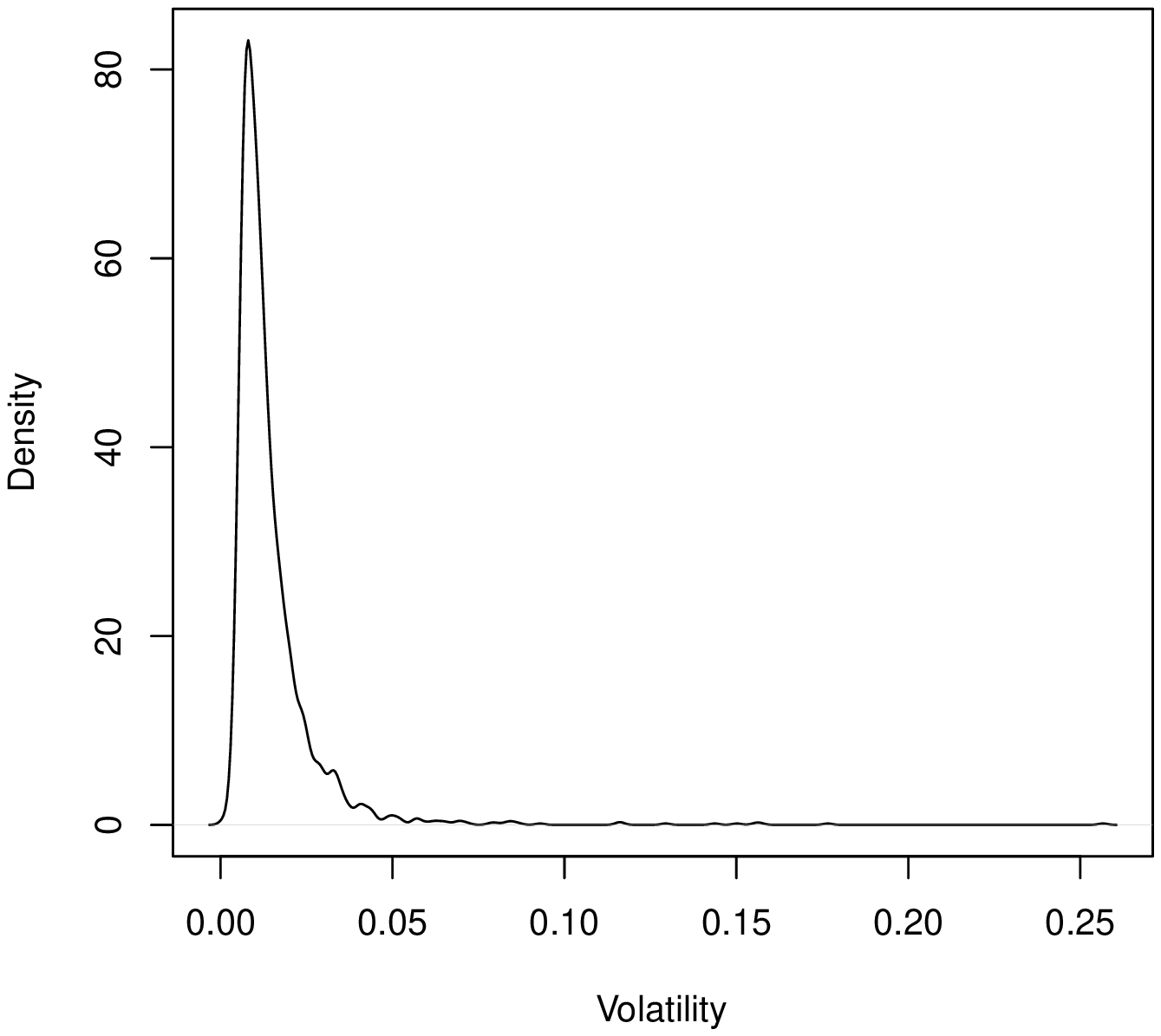}
\caption{Density of volatility for the universe of 2000 most liquid stocks (see Section \ref{sub.why}).
}
\label{Fig3}
\end{figure}

\begin{figure}[ht]
\centering
\includegraphics[scale=1.0]{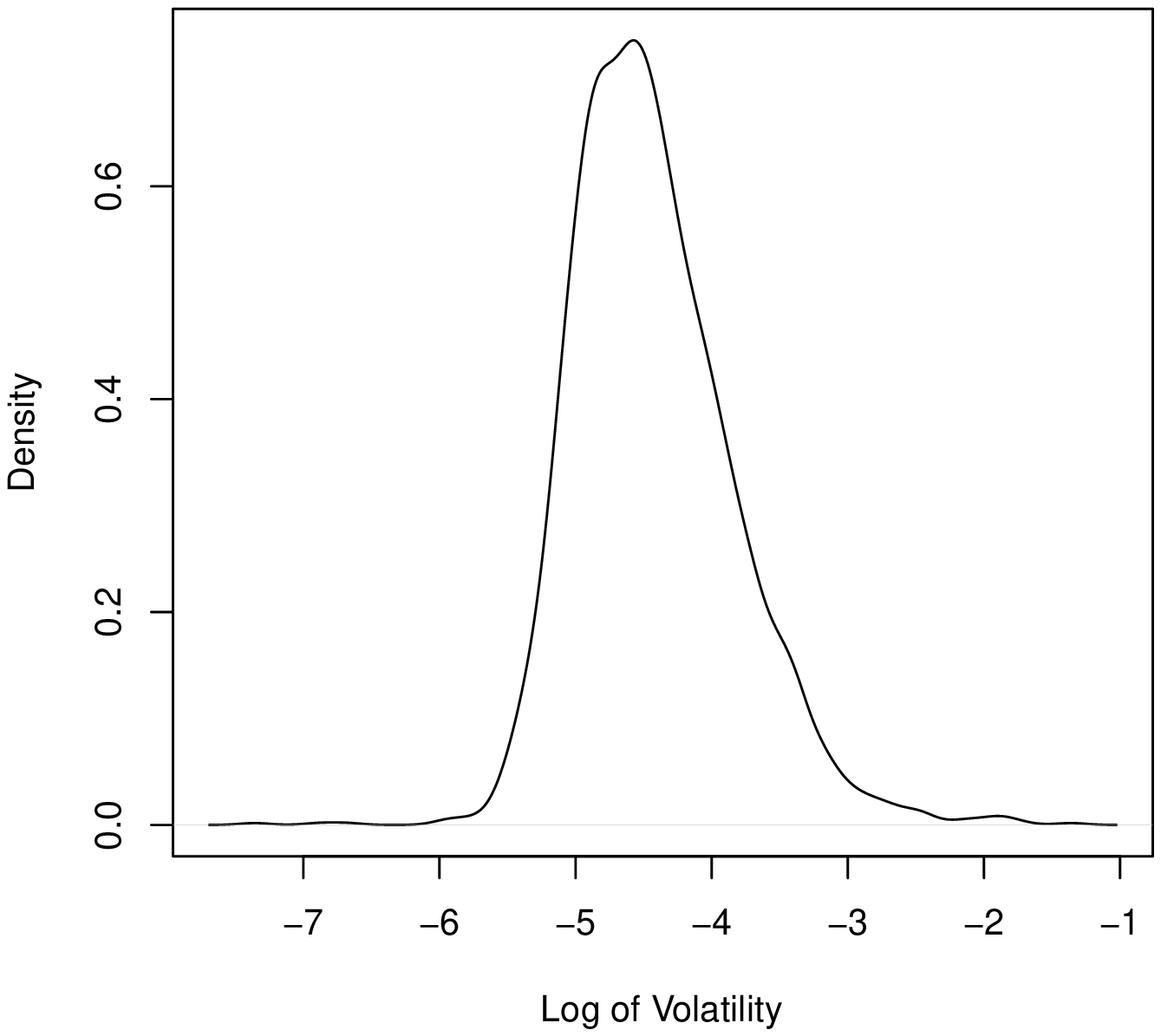}
\caption{Density of log-volatility for the universe of 2000 most liquid stocks (see Section \ref{sub.why}).
}
\label{Fig4}
\end{figure}

{}We can reasonably assume that there is no substantial correlation between ${\widetilde E}_i$ and $1/\sigma_i$. Then the deviations for the two terms in the parenthesis in (\ref{w.fano.1-factor}) can be estimated independently. The standard deviation of the term containing the sum in (\ref{w.fano.1-factor}) (approximately) is $\sqrt{2/N}~{\widetilde E}_*$. Conservatively, assuming that its actual value deviates by $5/\sqrt{2} \approx 3.54$ standard deviations in either direction, we can estimate the bound ${\widetilde \sigma}$ on $\sigma_i$ such that, for $\sigma_i < {\widetilde\sigma}$ it is unlikely (with roughly 3.54 standard deviations confidence level) that the term containing the sum in (\ref{w.fano.1-factor}) outweighs the second term in (\ref{w.fano.1-factor}) such that the total contribution of these terms is negative:
\begin{equation}\label{5SD}
 {\widetilde\sigma} \approx {\sqrt{N}\over 5}~\sigma_*
\end{equation}
And the number of such stocks typically is pretty small (compared with the number of stocks in the portfolio). To illustrate this, here is an example from data. We take the data for the universe of tickers as of Sep 6, 2014 that have historical pricing data on http://finance.yahoo.com (accessed on Sep 6, 2014) for the period Aug 1, 2008 through Sep 5, 2014.\footnote{\, The choice of this window is not critical here. We simply used data readily available to us.} We restrict this universe to include only U.S. listed common stocks and class shares (no OTCs, preferred shares, etc.) with BICS (Bloomberg Industry Classification System) sector, industry and sub-industry assignments as of Sep 6, 2014. The number of such tickers in our data is 3811. We then compute the 21-trading-day (i.e., 1-month) historical volatilities based on daily close-to-close returns for the most recent date in the data, Sep 5, 2014. One stock was not trading (zero volatility) in that 21-trading day period, so we are left with 3810 stocks with nonzero volatilities. These are our $\sigma_i$. The cross-sectional distribution of $\sigma_i$ is roughly log-normal, with a long tail at higher values (see Figures \ref{Fig1} and \ref{Fig2}). The summary of these 3810 values of $\sigma_i$ is as follows: Min = $0.64\times 10^{-3}$, 1st Quartile = $9.16\times 10^{-3}$, Median = 0.0137, Mean = 0.0185, 3rd Quartile = 0.02197, Max = 0.3252, SD (standard deviation) = 0.01706, MAD (mean absolute deviation) = $8.37\times 10^{-3}$. Further, we have $\sigma_* = 9.99\times 10^{-3}$, and the number of stocks in this universe with $\sigma_i \geq {\widetilde \sigma}$ is only 14. If we take 10 in the denominator\footnote{\, This corresponds to $10/\sqrt{2} \approx 7.07$ standard deviations (instead of $5/\sqrt{2} \approx 3.54$ -- see above).} instead of 5 in the definition (\ref{5SD}), we still only get 78 stocks with $\sigma_i \geq {\widetilde \sigma}$. If we restrict our stock universe to the top 2000 most liquid stocks by ADDV (average daily dollar volume, also computed based on the same 21-trading-day period), the results are similar (also see Figures \ref{Fig3} and \ref{Fig4}): Min = $0.64 \times 10^{-3}$, 1st Quartile = $7.86\times 10^{-3}$, Median = 0.0111, Mean = 0.01470, 3rd Quartile = 0.01678, Max = 0.2566, SD = 0.01408, MAD = $5.60\times 10^{-3}$. Further, we have $\sigma_* = 8.55\times 10^{-3}$, and the number of stocks in this universe with $\sigma_i \geq {\widetilde \sigma}$ is only 16. If we take 10 in the denominator instead of 5 in the definition of (\ref{5SD}), again we still only get 69 stocks with $\sigma_i \geq {\widetilde \sigma}$.

\subsection{Multifactor Risk Models}

{}In the preceding subsection we discuss a simple 1-factor model where the pair-wise correlations (after removing the ``market" mode) take two values, $\pm\rho$. (If we add back the ``market" mode with a uniform correlation $\rho_0$, then the pair-wise correlations in the resultant correlation matrix take two values, neither of which need be (but one of them can be) negative. Our discussion above can be generalized to multifactor models (with the ``market" mode removed). The math is more involved but the gist of it is captured by the 1-factor example we discuss above. Thus, we can reasonably assume that the returns $E_i$ are not significantly correlated with $\sigma_i$ or the factor loadings $\Omega_{iA}$, so that in optimizing the Fano ratio (as compared with the Sharpe ratio) the expected returns effectively get shifted by a positive additive contribution for most stocks, excepting large volatility stocks. As above, this results in fewer weights violating the bounds (\ref{bounds}) and the portfolio is also more diversified.

\section{Long-Short Portfolios}\label{sec.3}

{}Above we discuss long-only portfolios. What about long-short portfolios? To maximize the Fano ratio, we need to maximize the objective function (\ref{g1}). The modulus in (\ref{g1}) complicates things. First, its derivative is well-defined for $w_i\neq 0$ and for the subset $J = \{i|w_i \neq 0\}$ the maximization of $g$ in (\ref{g1}) is equivalent to $\partial g/\partial w_i = 0$, $i\in J$. For the sake of simplicity,\footnote{\, Here we will not delve into the $w_i = 0$ (and other important) subtleties. Such subtleties arise, e.g., in the case of mean-variance optimization with linear costs. For a recent discussion, see, e.g., \cite{MeanRev}. For a partial list of related literature, see, e.g., \cite{Adcock}, \cite{Best}, \cite{Cadenillas}, \cite{Janecek}, \cite{Kellerer}, \cite{Lobo}, \cite{Mokkhavesa}, \cite{Patel}, \cite{Shreve}, and references therein.} let us assume that all $w_i\neq 0$. Then we have all the same formulas as above for the long-only portfolio (without any bounds as $w_i$ need no longer be nonnegative) except that $\nu_i$ is replaced by $\chi_i = \mbox{sign}(w_i)$. So the analog of (\ref{cc1}) now must be solved iteratively. However, here we will not delve into solving this problem (or its subtleties) as there is a more prosaic issue to address.

{}Ignoring the aforesaid subtleties, the equation we would need to solve iteratively reads (see (\ref{cc1}) and the subsequent equations for definitions of $a$ and $b$)
\begin{equation}\label{w.chi}
 w_i = \sum_{j=1}^N C^{-1}_{ij} \left[a~E_j + b~\chi_j\right]
\end{equation}
where $a$ and $b$ also depend on $\chi_i$. However, it is not this dependence that is problematic. Instead, it is the presence of signs, i.e., $\chi_i$, in (\ref{w.chi}). Signs are highly unstable (they ``flip-flop" a lot, especially for shorter horizons). To illustrate this, let us simplify things and consider the case of a diagonal covariance matrix $C_{ij} = \sigma_i^2~\delta_{ij}$. Then ${\widehat\eta}_i = E_i /\sigma_i^2$ and we can set $\chi_i = \mbox{sign}(E_i)$. So, for a small $E_i$ (e.g., compared with its historical standard deviation or some suitable multiple thereof), if its sign flips (but the absolute value remains small), we can have a 100\% opposite contribution from $\chi_i$ into (\ref{w.chi}). This is the root-cause of the aforesaid instability, which also persists even for non-diagonal $C_{ij}$ (in which case things are simply messier). We can think about this as follows. The weights (\ref{w.chi}) effectively are the same as linearly combining two strategies. One is based on optimizing the Sharpe ratio for the expected returns $E_i$. The other is based on optimizing the Sharpe ratio for {\em binary}\footnote{\, For the sake of simplicity, assuming, as above, that all $w_i\neq 0$, that is. If some $w_i=0$, then the corresponding returns are not binary but trinary (with at most a small number of null returns). However, this does not alter the above conclusion relating to the instability of the signs $\chi_i$.} expected returns $\chi_i = \pm 1$. It should come as no surprise to quant traders that the second strategy is suboptimal. E.g., if we take $\mbox{sign}(E_i)$ instead of $E_i$ as binary expected returns, this strategy underperforms the strategy based on optimizing $E_i$. This is because forecasting just the direction and not the magnitude of the expected returns provides only partial information. Linearly combining such a suboptimal strategy with the strategy based on optimizing the Sharpe ratio then also is suboptimal.

{}Can we fix this? We can smooth out the sign in $\chi_i = \mbox{sign}(w_i)$. One way to do this is to replace it by, e.g., a hyperbolic tangent: $\chi_i = \tanh(w_i/\Delta_i)$, where $\Delta_i$ are some parameters. In the limit $\Delta_i\rightarrow 0$ we recover $\chi_i = \mbox{sign}(w_i)$. Introducing $N$ new parameters $\Delta_i$ can be unappealing as they can easily turn out to be out-of-sample unstable. We can mitigate this, at least to a degree, by taking uniform $\Delta_i \equiv \Delta$ (however, we will relax this below). We then have
\begin{equation}\label{w.chi.1}
 w_i = \sum_{j=1}^N C^{-1}_{ij} \left[a~E_j + b~\tanh(w_j/\Delta)\right]
\end{equation}
We can solve this equation, e.g., by linearizing the hyperbolic tangent, which {\em formally} amounts to the limit where $\Delta\rightarrow \infty$, $b\rightarrow \infty$, and ${\widetilde b} = b/\Delta$ is kept finite:
\begin{equation}\label{w.chi.2}
 w_i = \sum_{j=1}^N C^{-1}_{ij} \left[a~E_j + {\widetilde b}~w_j\right]
\end{equation}
A formal solution\footnote{\, We, yet again, use the adjective ``formal" as $a$ and ${\widetilde b}$ a priori are undetermined (see below).} reads
\begin{eqnarray}\label{w.chi.3}
 && w_i = a \sum_{j=1}^N {\widetilde C}^{-1}_{ij}~ E_j\\
 && {\widetilde C}_{ij} = C_{ij} - {\widetilde b}~\delta_{ij}
\end{eqnarray}
The overall normalization parameter $a$ is fixed by requiring the normalization condition (\ref{norm}). However, the parameter ${\widetilde b}$ is a priori undetermined. Since we have departed from optimizing the Fano ratio, it is no longer evident what ${\widetilde b}$ should be. Instead of trying to fix it ``theoretically", we can take a pragmatic approach and treat ${\widetilde b}$ as a free parameter. For ${\widetilde b} = 0$ we are simply optimizing the Sharpe ratio. For ${\widetilde b} \neq 0$, we are optimizing the Sharpe ratio but with a modified covariance matrix ${\widetilde C}_{ij}$, whose off-diagonal elements are the same as those of $C_{ij}$, but the diagonal elements (variances) are shifted: they can be increased (${\widetilde b} < 0$) or decreased (${\widetilde b} > 0$).

{}In this regard, it is instructive to consider the case of nonuniform $\Delta_i$. In this case we still have (\ref{w.chi.3}), where now
\begin{equation}
 {\widetilde C}_{ij} = C_{ij} - {\widetilde b}_i~\delta_{ij}
\end{equation}
and ${\widetilde b}_i = b/\Delta_i$. Let us consider a factor model of the form (\ref{factor}). If we set ${\widetilde b}_i = \theta~\xi_i^2$, where $\xi_i^2$ are the specific variances and $\theta$ is a parameter, then for $\theta = 1$ the matrix ${\widetilde C}_{ij}$ is singular. We can invert it in the $\theta\uparrow 1$ limit (in this limit the normalization $a$ goes to 0 such that $w_i$ are actually finite) and the result is that, up to an overall normalization factor (fixed via (\ref{norm})), the weights $w_i$ are given by $\epsilon_i/\xi_i^2$, where $\epsilon_i$ are the residuals of a cross-sectional regression of $E_i$ over the factor loadings $\Omega_{iA}$ with the regression weights $z_i = 1/\xi_i^2$ and no intercept\footnote{\, Unless the intercept is already subsumed in the factor loadings matrix $\Omega_{iA}$, that is.} \cite{MeanRev}. Equivalently, ${\widetilde \epsilon}_i = \epsilon_i/\xi_i$ are the residuals of a cross-sectional regression of $E_i/\xi_i$ over the matrix $\Omega_{iA}/\xi_i$ with unit regression weights (and no intercept -- see above). So, here we are interpolating between optimizing the Sharpe ratio and a (weighted) regression.

{}Formally, we can view (\ref{w.chi.3}) as an infinite series (here $a_p = a~{\widetilde b}^{p-1}$):
\begin{eqnarray}
 && w_i = \sum_{p=1}^{\infty} a_p~\sum_{j = 1}^N C^{-1}_{ij}~E^{(p)}_j\\
 && E^{(p+1)}_i = \sum_{j=1}^N C^{-1}_{ij}~E^{(p)}_j\label{Ep}\\
 && E^{(1)}_i = E_i
\end{eqnarray}
I.e., this is a combination of ``once-optimized", ``twice-optimized", ``trice-optimized", \dots, strategies. In fact, we can simply forget about how we got this result (which was in an ad hoc and handwaving fashion -- however, see below) and take a truncated series
\begin{equation}\label{w.chi.4}
 w_i = \sum_{p=1}^{n_{opt}} a_p~\sum_{j = 1}^N C^{-1}_{ij}~E^{(p)}_j = a~\sum_{j = 1}^N C^{-1}_{ij}~{\widehat E}_j
\end{equation}
where ${\widehat E}_i = \sum_{p=1}^{n_{opt}} {\widetilde b}^{p-1}~E^{(p)}_i$. Only one of the $n_{opt}$ coefficients $a_p$ is fixed by the normalization condition (\ref{norm}), i.e., we can fix $a_1 = a$. As mentioned above, a priori there is no guiding principle for fixing the ${\widetilde b}$ parameter.\footnote{\, More generally, we can depart from $a_p = a~{\widetilde b}^{p-1}$ and treat the coefficients $a_p$ as independent. Then we can datamine $a_{p>1}$ and see if they are stable out-of-sample. We will not do this here.} However, we can require that the coefficients $a_p$ have the proper scaling properties under $E_i\rightarrow \zeta E_i$ and $C_{ij}\rightarrow \lambda C_{ij}$, where $\zeta > 0$ and $\lambda > 0$ (so that $w_i$ are invariant under such rescalings):
\begin{eqnarray}
 && a_p \rightarrow \zeta^{-1}~a_p\\
 && a_p \rightarrow \lambda^p~a_p
\end{eqnarray}
This implies that ${\widetilde b}$ is invariant under $E_i\rightarrow \zeta E_i$, and we have ${\widetilde b} \rightarrow\lambda {\widetilde b}$ under $C_{ij}\rightarrow \lambda C_{ij}$. Consider ${\widetilde b}$ of the following form:
\begin{equation}
 {\widetilde b} = {\widehat b}~h = {\widehat b}~\sqrt{{\sum_{i,j=1}^N} C_{ij}^{-1}~E_i~E_j \over {\sum_{i,j=1}^N} C_{ij}^{-1}~E^{(2)}_i~E^{(2)}_j}
\end{equation}
Then ${\widehat b}$ is invariant under both the $\zeta$ and $\lambda$ rescalings. We can therefore treat ${\widehat b}$ as a purely numerical coefficient. For instance, for $n_{opt} = 2$ we have
\begin{equation}
 w_i = a\left[\sum_{j = 1}^N C^{-1}_{ij}~E_j + {\widehat b}~h~\sum_{j = 1}^N C^{-1}_{ij}~E^{(2)}_j\right]
\end{equation}
I.e., we are combining the ``once-optimized" and ``twice-optimized" strategies with the relative coefficient controlled by ${\widehat b}$. We discuss a backtest of this strategy below.

\subsection{Bells and Whistles}\label{sub.bells}

{}While our $w_i$ in (\ref{w.chi.4}) are roughly dollar-neutral (due to the presence of the ``market" mode in $C_{ij}$, which a priori need not be removed for long-short portfolios), they are not exactly dollar-neutral. We may wish our long-short portfolio to be exactly dollar-neutral (e.g., due to risk management/compliance requirements, etc.):
\begin{equation}
 \sum_{i=1}^N w_i = 0
\end{equation}
More generally, we may wish to impose more than one linear homogeneous constraints
\begin{equation}\label{lin.1}
 \sum_{i=1}^N G_{i\alpha}~w_i = 0,~~~\alpha = 1,\dots,m
\end{equation}
where the columns of the $N\times m$ matrix $G_{i\alpha}$ are linearly independent. Such constraints are readily incorporated in the optimization problem by ``padding" the factor loadings matrix with the extra $m$ columns: ${\widetilde \Omega}_{i{\widetilde A}} = (\Omega_{iA}, G_{i\alpha})$, where the index ${\widetilde A} = (A, \alpha)\in H$ now takes ${\widetilde K} = |H| = K + m$ values ($H = \{{\widetilde A}\}$). We then have (see, e.g., \cite{MeanRev})
\begin{eqnarray}
 &&C^{-1}_{ij} = {1\over\xi_i^2}~\delta_{ij} - \sum_{{\widetilde A},{\widetilde B}\in H} {{\widetilde \Omega}_{i{\widetilde A}}\over\xi_i^2}~{\widetilde Q}^{-1}_{{\widetilde A}{\widetilde B}}~{{\widetilde \Omega}_{j{\widetilde B}}\over\xi_j^2}\\
 &&{\widetilde Q}_{{\widetilde A}{\widetilde B}} = \varphi_{{\widetilde A}{\widetilde B}} + \sum_{i\in J} {1\over\xi_i^2}~{\widetilde \Omega}_{i{\widetilde A}}~{\widetilde \Omega}_{i{\widetilde B}}\\
 &&\varphi_{AB} = \phi_{AB}^{-1},~~~A,B=1,\dots,K\\
 &&\varphi_{A\alpha} = 0,~~~A=1,\dots,K,~\alpha=1,\dots,m\label{varphi.2}\\
 &&\varphi_{\alpha\beta} = 0,~~~\alpha,\beta=1,\dots,m\label{varphi.3}
\end{eqnarray}
The matrix $C^{-1}_{ij}$ has the following property:
\begin{equation}
 \sum_{j=1}^N C^{-1}_{ij}~{\widetilde \Omega}_{j{\widetilde C}} = \sum_{{\widetilde A},{\widetilde B}\in H} {{\widetilde \Omega}_{i{\widetilde A}}\over\xi_i^2}~{\widetilde Q}^{-1}_{{\widetilde A}{\widetilde B}}~\varphi_{{\widetilde B}{\widetilde C}}
\end{equation}
which (together with (\ref{varphi.2}) and (\ref{varphi.3})) in turn implies that
\begin{equation}
 \sum_{j=1}^N C^{-1}_{ij}~G_{j\alpha} \equiv 0,~~~\alpha=1,\dots,m
\end{equation}
This results in a solution (\ref{w.chi.4}) satisfying the linear constrains (\ref{lin.1}). In practice, to minimize noise in the factor model covariance matrix, the factor loadings $\Omega_{iA}$ should be chosen orthogonal to the matrix $G_{i\alpha}$ \cite{MeanRev}:
\begin{equation}\label{ortho.G}
 \sum_{i=1}^N \Omega_{iA}~G_{i\alpha}\equiv 0,~~~A=1,\dots,K,~\alpha=1,\dots,m
\end{equation}
This is not required for the above ``padding" trick, which works irrespective of (\ref{ortho.G}).\footnote{\, Also, the ``padding" is needed only in $C_{ij}$ on the r.h.s. of (\ref{w.chi.4}), not in the definitions (\ref{Ep}).}

{}Another consideration is that in practice one often needs to impose upper and lower bounds on $w_i$:
\begin{equation}\label{bounds.m}
 w_i^-\leq w_i \leq w_i^+
\end{equation}
See \cite{MeanRev} for a practically-oriented discussion. Assuming $w_i^- < 0$ and $w_i^+ > 0$, we can readily incorporate such bounds using the algorithm given in \cite{MeanRev} for which the source code is given in \cite{Het}. The bounds (\ref{bounds.m}) are simply imposed in optimizing the Sharpe ratio with the ``effective" expected returns ${\widehat E}_i$ on the r.h.s. of (\ref{w.chi.4}) (but no bounds are imposed in (\ref{Ep})).

{}Finally, for our backtesting purposes below, here we discuss how to include trading costs. Including nonlinear impact complicates the problem and is unnecessary for our purposes here. However, we can include linear trading costs. Below we will consider purely intraday strategies where the positions are established just once at the open and are liquidated just once at the close of the same trading day. For the stock labeled by $i$, let the linear trading cost per {\em dollar} traded be $\tau_i$. Then including such costs in the case of optimizing the Sharpe ratio with the expected returns $E_i$ amounts to replacing the expected return for the portfolio (\ref{E}) by\footnote{\, This is the expected return of the portfolio once it is established. In computing the P\&L, we must take into account not only the establishing costs, but also the liquidating costs (so the total costs subtracted from the P\&L are approximately double the establishing costs).}
\begin{equation}
 E = \sum_{i=1}^N \left[E_i~w_i - \tau_i~|w_i|\right]
\end{equation}
A complete algorithm for including linear trading cost in mean-variance optimization is given in, e.g., \cite{MeanRev}. However, for our purposes here the following simple ``hack" suffices. We can define the effective return
\begin{equation}\label{E.tc}
 E_i^{eff} = \mbox{sign}(E_i)~\mbox{max}(|E_i| - \tau_i, 0)
\end{equation}
and simply set
\begin{equation}
 E = \sum_{i=1}^N E_i^{eff}~w_i
\end{equation}
I.e., if the magnitude for the expected return for a given stock is less than the expected cost to be incurred, we set the expected return to zero, otherwise we reduce said magnitude by said cost. This way we can avoid a nontrivial iterative procedure (see \cite{MeanRev}), albeit we emphasize that this solution is only an approximation to the optimal solution. However, here we are already employing other approximations, so this way of treating linear trading costs is well-justified.\footnote{\, For ``multiply-optimized" strategies in (\ref{w.chi.4}), it may make sense to use more sophisticated approximations. For the sake of simplicity and not to overcomplicate things, we will use (\ref{E.tc}) here.}

{}So, what should we use as $\tau_i$ in (\ref{E.tc})? The model of \cite{Almgren} is reasonable for our purposes here. Let $H_i$ be the {\em dollar} amount traded for the stock labeled by $i$. Then for the linear trading costs we have
\begin{equation}
 T_i = \zeta~\sigma_i~{|H_i|\over A_i}
\end{equation}
where $\sigma_i$ is the historical volatility, $A_i$ is the average daily dollar volume (ADDV), and $\zeta$ is an overall normalization constant we need to fix. However, above we work with weights $w_i$, not traded dollar amounts $H_i$. In our case of a purely intraday trading strategy discussed above, they are related simply via $H_i = I~w_i$, where $I$ is the total investment level (i.e., the total absolute dollar holdings of the portfolio after establishing it). Therefore, we have (note that $T_i = \tau_i~|H_i| = \tau_i~I~|w_i|$)
\begin{equation}
 \tau_i = \zeta~{\sigma_i~\over A_i}
\end{equation}
We will fix the overall normalization $\zeta$ via the following heuristic. We will (conservatively) assume that the average linear trading cost per dollar traded is 10 bps (1 bps = 1 basis point = 1/100 of 1\%),\footnote{\, This amounts to assuming that, to establish an equally-weighted portfolio, it costs 10 bps.} i.e., $\mbox{mean}(\tau_i) = 10^{-3}$ and $\zeta = 10^{-3} / \mbox{mean}(\sigma_i / A_i)$.

\subsection{Backtests}\label{sub.back}

{}Here we discuss some backtests. We wish to see how ``multiply-optimized" strategies (\ref{w.chi.4}) for dollar-neutral intraday models compare with optimizing the Sharpe ratio (i.e., a ``singly-optimized" strategy). For this comparison, we run our backtests as in \cite{Het}. For our $C_{ij}$ (in all cases) we use heterotic risk models of \cite{Het}. The historical data we use in our backtests here is the same as in \cite{Het} and is described in detail in Subsections 6.2 and 6.3 thereof. The trading universe selection is described in Subsection 6.2 of \cite{Het}. We assume that the portfolio is established at the open with fills at the open prices; and ii) it is liquidated at the close on the same day -- so this is a purely intraday strategy -- with fills at the close prices. We include the transaction costs as discussed in Subsection \ref{sub.bells} hereof. Furthermore, we include strict trading bounds (which in this case are the same as position bounds)
\begin{equation}
 |H_i| \leq 0.01~A_i
\end{equation}
We further impose strict dollar-neutrality on the portfolio, so that
\begin{equation}
 \sum_{i=1}^N H_i = 0
\end{equation}
The total investment level in our backtests here is $I$ = \$20M (i.e., \$10M long and \$10M short), same as in \cite{Het}. For the Sharpe ratio optimization with bounds we use the R function {\tt{\small bopt.calc.opt()}} in Appendix C of \cite{Het}. We use ${\widehat b}=1$ (see above) in ``multiply-optimized" strategies. The backtest results are summarized in Table \ref{table1}, which shows that the $n_{opt} = 2$ strategy outperforms the $n_{opt} = 1$ strategy (which is simply optimizing the Sharpe ratio). However, for higher $n_{opt}$ it appears that we get -- quite literally -- ``diminishing returns".

\section{Concluding Remarks}\label{sec.4}

{}For long-only portfolios optimizing the Fano ratio effectively amounts to shifting the expected returns by a positive amount via (\ref{E-eff-long}), which results in fewer stocks being excluded from the portfolio due to the bounds (\ref{bounds}) (including some stocks with negative expected returns $E_i$, for which effective returns ${\widetilde E}_i$ can be positive), i.e., in a more diversified portfolio\footnote{\, And also less skewed portfolio.} (compared with optimizing the Sharpe ratio).

{}However, for long-short portfolios this is a non-issue to begin with: the weights need not be nonnegative. As we discuss above, optimizing the Fano ratio in this case would be suboptimal. However, the Fano ratio optimization inspires considering modifications of optimizing the Sharpe ratio, such as (\ref{w.chi.1}) and (\ref{w.chi.2}). In this regard, the following comment is in order. Linearizing the hyperbolic tangent in (\ref{w.chi.1}) amounts to completely removing the sign ``flip-flopping" issue discussed in Section \ref{sec.3}, which (to a lesser degree) is present even when we replace the sign in (\ref{w.chi.1}) by the hyperbolic tangent in (\ref{w.chi.2}). The further reduction via (\ref{w.chi.4}) essentially amounts to simply combining multiple different alphas -- even though here alphas are of a specific (``multiply-optimized") form. However, more generally, combining multiple (even a large number of) different alphas yields higher returns and Sharpe ratios and lower turnover and higher cents-per-share (see, e.g., \cite{Billion}).

{}Finally, let us mention that the Fano ratio arises in the context of statistical industry classifications via clustering techniques \cite{StatIC}. One question for clustering in the context of quant trading is what to cluster? Clustering returns is suboptimal. Naively, clustering normalized returns $E_i/\sigma_i$ appears to be reasonable. However, as was argued and supported via backtests in \cite{StatIC}, clustering $E_i/\sigma_i^2$ -- i.e., the corresponding Fano ratios -- is the optimal choice. Thus, clustering $E_i/\sigma_i$ groups together stocks that are (to varying degrees) highly correlated in-sample. However, there is no guarantee that they will remain as highly correlated out-of-sample. Intuitively, it is evident that higher volatility stocks are more likely to get uncorrelated with their respective clusters. This is essentially why suppressing by another factor of $\sigma_i$ in the Fano ratio $E_i/\sigma_i^2$ (as compared with $E_i/\sigma_i$) leads to better performance: inter alia, it suppresses contributions of those volatile stocks into the corresponding cluster centers \cite{StatIC}.

%\clearpage
%\newpage

\begin{table}[ht]
\caption{Simulation results for the ``multiply-optimized" strategies (defined in Eq. (\ref{w.chi.4}) in Section \ref{sec.3}) with bounds and costs. ROC = Return-on-Capital, SR = annualized Sharpe Ratio, CPS = cents-per-share. Note that $n_{opt}=1$ corresponds to vanilla optimization of the Sharpe ratio (i.e., a ``singly-optimized" strategy).} % title of Table
%\centering % used for centering table
\begin{tabular}{l l l l} % centered columns (4 columns)
\\
\hline\hline %inserts double horizontal lines
$n_{opt}$ & ROC & SR & CPS\\[0.5ex] % inserts table
%heading
\hline % inserts single horizontal line
1 & 35.37\% & 13.65 & 1.74\\
2 & 36.62\% & 15.43 & 2.02\\
3 & 34.00\% & 15.39 & 2.06\\
4 & 26.38\% & 11.80 & 1.69\\
5 & 17.00\% & 7.27 & 1.09\\[1ex] % [1ex] adds vertical space
\hline %inserts single line
\end{tabular}
\label{table1} % is used to refer this table in the text
\end{table}

\clearpage
\newpage

\clearpage
\newpage

\clearpage
\newpage

\clearpage
\newpage

\end{document}